\documentclass[aps,pre,reprint,superscriptaddress,amsmath,amssymb,showpacs]{revtex4-1}
\usepackage{graphicx}
\usepackage{bm}
\usepackage{color}
\usepackage{lipsum}
\usepackage{soul}
\usepackage{mathtools}
\usepackage{bbm}
\usepackage{url}

\begin{document}

\setcounter{page}{1} 

\title{Thermodynamic constraints on the nonequilibrium response of one-dimensional diffusions }
\author{Qi Gao}
\affiliation{Department of Physics, University of Michigan, Ann Arbor, Michigan, 48109, USA}
\author{Hyun-Myung Chun}
\affiliation{Department of Biophysics, University of Michigan, Ann Arbor, Michigan 48109, USA}
\author{Jordan M. Horowitz}
\affiliation{Department of Physics, University of Michigan, Ann Arbor, Michigan 48109, USA}
\affiliation{Department of Biophysics, University of Michigan, Ann Arbor, Michigan, 48109, USA}
\affiliation{Center for the Study of Complex Systems, University of Michigan, Ann Arbor, Michigan 48104, USA}

\date{\today}
\begin{abstract}
We analyze the static response to perturbations of nonequilibrium steady states that can be modeled as one-dimensional diffusions on the circle. We demonstrate that an arbitrary perturbation can be broken up into a combination of three specific classes of perturbations that can be fruitfully addressed individually. For each class, we derive a simple formula that quantitatively characterizes the response in terms of the strength of nonequilibrium driving valid arbitrarily far from equilibrium.
\end{abstract}

\maketitle 

\emph{Introduction.}\ Linear response theory developed as a tool to rationalize the response of equilibrium systems to external perturbations and internal fluctuations. Its central organizing prediction, the fluctuation-dissipation theorem (FDT)~\cite{Kubo}, characterizes such response in terms of experimentally measurable equilibrium correlation functions.
This result has been immensely useful, helping to form the statistical-mechanical foundation of hydrodynamics, establishing Green-Kubo relations~\cite{Kadanoff1963,Forster}, as well as providing the theoretical scaffolding for light scattering and microrheology experiments~\cite{chaikin_lubensky_1995}.

Motivated by these early successes, it is now customary to probe a system's behavior, no matter how far from equilibrium, in terms of responses to perturbations and correlation functions. Examples can be found in studies of active matter \cite{Fily2012,Bialke2013,Martin14380,Mizuno370} as well as in analyses of biological function~\cite{Sato2003,Murugan2011,Qian2012,Yan2013,Hartich2015,Estrada2016}. However, without the simplicity of the equilibrium FDT as a guiding principle, disparate analysis methods have emerged. One approach has been to re-establish the connection between response and correlation functions around nonequilibrium steady states~\cite{Agarwal1972}.
While the correlation functions require detailed knowledge of the system's microscopic dynamics, recent theoretical insights from stochastic thermodynamics have provided them with crisp physical interpretations in terms of stochastic entropy production and dynamical activity~\cite{Prost2009,Baiesi2009,Seifert2010}.
A complementary approach has been to characterize violations of the equilibrium-version of the FDT, either through the introduction of effective temperatures \cite{BenIsaac2011,Cugliandolo_2011,Dieterich2015} or for Brownian particles by connecting the violation directly to the steady state entropy production via the Harada-Sasa equality \cite{PhysRevLett.95.130602,Toyabe2010}. 

In the tradition of studying violations of the FDT, one of us recently demonstrated that the magnitude of the response to an external perturbation can be quantitatively constrained by the degree of nonequilibrium driving~\cite{PhysRevX.10.011066}.
These predictions were limited to static (or zero-frequency) response in nonequilibrium steady states that could be modeled as discrete continuous-time Markov jump processes with a finite number of states. In this article, we expand this framework to the static response of nonequilibrium steady states described by one-dimensional diffusion processes with periodic boundary conditions. This class of systems not only encompasses a variety of experimental situations, such as a driven colloidal particle in a viscous fluid~\cite{Speck2007,Joubaud2008,GomezSolano2009}, but is also analytically tractable, which has made it a paradigmatic theoretical model within stochastic thermodynamics~\cite{Seifert2012}.

Our main contribution is to unravel an arbitrary perturbation of a diffusive steady state into a linear combination of three classes of perturbations that can be individually analyzed.
For each class we prove an equality or inequality that quantifies how thermodynamics and nonequilibrium driving constrain the response.

\emph{Setup.}\ Our focus is a single periodic degree of freedom $x$ that evolves diffusively on a circle of length $L$. The dynamics are completely characterized by the probability density $\rho(x,t)$ as a function of time $t$ and position $x$ whose evolution is governed by the generic Fokker-Planck equation~\cite{Gardiner},
\begin{equation}
\begin{split}
\partial_t \rho(x,t)&=-\partial_x [A(x)\rho(x,t)]+\partial_ x [B(x)\partial_x\rho(x,t)]\\
&\equiv \hat{\mathcal L} \rho(x,t),
\end{split}
\label{FPE}
\end{equation}
with periodic functions $A(x)$ and $B(x)$.
Equation~\eqref{FPE} has a unique steady state distribution $\pi(x)$, given as the periodic solution of $\hat{\mathcal L}\pi(x)=0$.
In general, $\pi(x)$ represents a nonequilibrium steady state.
However, when the functions $A(x)$ and $B(x)$ satisfy the potential condition $\int_0^L A(z)/B(z)dz =0$, the dynamics are detailed balanced and the resulting steady state describes an equilibrium situation $\pi^{\rm eq}(x)\propto e^{\psi(x)}$ with conservative potential $\psi(x)=\int_0^x A(z)/B(z)dz $~\cite{Gardiner}. Indeed, the magnitude of the breaking of the potential condition can be identified with the thermodynamic force ${\mathcal F}=\int_0^LA(z)/B(z)dz $ driving the system away from equilibrium, when the dynamics are thermodynamically consistent~\cite{Polettini2016,Polettini2016b}.

\emph{Parametrizing steady-state response.}\ Our aim is to characterize how steady state averages of observables  $\langle Q\rangle = \int_0^L Q(z) \pi(z)dz$ change in response to variations in $A(x)$ and $B(x)$.
Our main contribution here is to recognize that it is useful to parametrize changes in the dynamics with a constant $f$ and two periodic functions $\mu(x)$ and $U(x)$ via $A(x)=\mu(x)[-U'(x)+f]$ and $B(x)=\mu(x)$:
\begin{equation}\label{eq:paraDecomp}
\hat{\mathcal L}\rho=-\partial_x{\mu(x) [- U'(x)+f]\rho} + \partial_x [\mu(x)\partial_x \rho].
\end{equation}
We were led to this parametrization by first discretizing the diffusion process and then 
comparing the result to the decomposition introduced previously in \cite{PhysRevX.10.011066} for discrete Markov jump processes.
This mapping then suggested that derivatives with respect to $\mu$, $U$, and $f$ could have interesting thermodynamic limits.  
While the analysis here is completely self-contained given the definitions in \eqref{eq:paraDecomp}, we do include for reference the discretization mapping in~\cite{SM}.

More general perturbations in $A(x)$ and $B(x)$ can then be built up as linear combinations of changes in $\mu$, $U$ and $f$. Indeed, if we perturb the dynamics by making infinitesimal changes $A(x)\to A(x)+\delta A(x)$ and $B(x)\to B(x)+\delta B(x)$, then changes in our parameters can be conveniently expressed in terms of $\Delta(x)=[\delta A(x) B(x)-\delta B(x) A(x)]/B(x)^2$ as~\cite{SM}
\begin{align}
&\delta \mu(x) =\delta B(x), \\
&\delta U(x) = -\int_0^x \Delta (z) \ dz + \frac{x}{L}\int_0^L \Delta (z)\ dz + \delta U(0), \\
&\delta f = \frac{1}{L}\int_0^L \Delta(z)\ dz,
\end{align}
where $\delta U(0)$ is an undetermined constant, which does not affect the predictions.

While our parametrization is a mathematical convenience, the notation here is meant to bring to mind the equation of motion of a colloidal particle in a viscous fluid at (dimensionless) temperature $k_B T=1$ with spatially-dependent mobility $\mu(x)$ moving in an energy landscape $U(x)$ driven by a constant nonconservative mechanical force $f$. We will rely on this analogy for intuition, and often use this terminology. However, we stress that this is only a mathematical equivalence and our analysis is not restricted to a single overdamped particle, but applies to any physical system that can be accurately modeled as a one-dimensional diffusion.
Indeed, any model specified by $A(x)$ and $B(x)$ can be mapped to our parametrization.
Moreover, our decomposition captures the most general separation of the dynamics into a conservative contribution $U(x)$ and a nonconservative contribution $f$.
This highlights the fact that the only way to break the potential condition is the inclusion of a force with a constant contribution $f$, with the resulting thermodynamic force ${\mathcal F}=\int_0^L A(z)/B(z)dz =f L$. Thermodynamic equilibrium is then characterized by $f=\frac{1}{L}\int_0^L A(z)/B(z)dz=0$, in which case the steady-state distribution takes the Gibbs form $\pi^{\rm eq}(x)\propto e^{-U(x)}$ in terms of the (dimensionless) energy landscape. 
From this point of view, perturbations of $A$ and $B$ usually amount to affecting only $U$ or $f$~\cite{Baiesi2013}. We find here that by allowing for perturbations in $\mu$ in our theoretical analysis, we are able to unravel simple limits on response, even if perturbations that end up only affecting $\mu$ in experimental settings may not be common. Our main predictions are then a series of equalities and inequalities for the steady state averages of observables due to perturbations in our three functions $\mu$, $U$, and $f$. 

Our first prediction is an equality for the response of an arbitrary observable $Q$ to a coupled $U$ and $\mu$ perturbation,
\begin{equation}
\label{eq:Ebound}
\frac{\delta \langle Q\rangle}{\delta U(y)}+\frac{\delta\langle Q\rangle}{\delta \ln \mu(y)}=-\pi(y)[Q(y)-\langle Q\rangle].
\end{equation}

For $\mu$-perturbations, we derive an inequality on the ratio of the averages of two nonnegative observables $Q_1$ and $Q_2$  ($Q_1,Q_2\ge 0$),
\begin{align}
\label{eq:Bbound1}
\left|\int_a^b \frac{\delta \ln (\langle Q_1\rangle/\langle Q_2\rangle)}{\delta \ln \mu(z)}\ dz\right|\le \tanh( |{\mathcal F}|/4).
\end{align}
Note that the restriction to non-negative observables does not pose any serious limitation as we can always shift any observable by its minimum to create a non-negative one.

Last, we find that constraints on $f$ perturbations can most naturally be expressed as responses to the thermodynamic force ${\mathcal F}=fL$,
\begin{align}
\label{eq:fbound}
\left|\frac{\partial\ln(\langle Q_1\rangle/\langle Q_2\rangle)}{\partial {\mathcal F}}\right|\le 1.
\end{align}

By exploiting the freedom to choose the observables $Q_1$ and $Q_2$, we can arrive at bounds for a variety of quantities of interest. 
For example, the choice $Q_1(z; x) = \delta(z-x)$ and $Q_2=1$, gives bounds on the response of the steady-state density
\begin{align}
\label{eq:Bbound1b}
&\left|\int_a^b \frac{\delta \ln\pi(x)}{\delta \ln \mu(z)}\ dz\right|\le\tanh( |{\mathcal F}|/4),\ \\
\label{eq:Bbound1c}
&\left|\frac{\partial\ln\pi(x)}{\partial {\mathcal F}}\right|\le 1.
\end{align}

We obtain our results by differentiating the known analytic expression for the steady state distribution~\cite{Gardiner},
\begin{equation}
\begin{split}
\pi(x) & =\frac{e^{-U(x)+fx}}{\mathcal N}\Bigg[e^{-fL}\int_0^x e^{U(z)-fz-\ln\mu(z)}dz \\
& ~~~~~~~~~~~~~~~~~~~~~
+\int_x^L e^{U(z)-fz-\ln\mu(z)}dz\Bigg],
\end{split}
\end{equation}
with ${\mathcal N}$ a normalization constant, and then reasoning about the result. Derivations are presented in \cite{SM}.  Here, we examine and illustrate these formulas.

\emph{Equilibrium-like FDT.}\ At thermodynamic equilibrium (${\mathcal F}=0$), the response to perturbations in the energy landscape $U(x)$ is well characterized by the FDT in terms of equilibrium correlation functions. Imagine we perturb an equilibrium system by slightly altering an externally controllable parameter $\lambda$ that affects the energy as $U_\lambda(x)=U(x)-\lambda V(x)$, which defines the coordinate conjugate to the perturbation $V(x)$. The equilibrium FDT then predicts that the response of an arbitrary observable $Q(x)$ can be expressed as~\cite{MARCONI2008111}
\begin{equation}
\partial_\lambda \langle Q\rangle =  {\rm Cov}_{\rm eq}(Q,V),
\end{equation} 
in terms of the fluctuations via the equilibrium covariance ${\rm Cov}_{\rm eq}(Q,V)=\langle QV\rangle_{\rm eq}-\langle Q\rangle_{\rm eq}\langle V\rangle_{\rm eq}$.

Away from thermodynamic equilibrium (${\mathcal F}\neq 0$), the response to $U(x)$ perturbations is generally more challenging to characterize. However, when we combine changes in $U$ with $\mu$ as in  \eqref{eq:Ebound}, we find a response that is exactly equivalent to the response of an equilibrium Gibbs distribution to changes in $U$ alone.
We can exploit this observation by considering a perturbation that is equivalent to varying the energy and mobility in concert as 
$U_\lambda(x)=U(x)-\lambda V(x)$ and $\mu_\lambda(x)=\mu(x)[1-\lambda V(x)]$.
In this case, the response is
\begin{align}\label{eq:FDT1}
\partial_\lambda \langle Q\rangle
&=-\int_0^L V(z)\left [\frac{\delta\langle Q\rangle}{\delta U(z)}+\frac{\delta\langle Q\rangle}{\delta \ln\mu(z)}\right]\ dz.
\end{align}
A direct application of \eqref{eq:Ebound} then allows us to interpret the result as a simple FDT-like expression 
\begin{equation}\label{eq:FDT2}
\partial_\lambda \langle Q\rangle = {\rm Cov}(Q,V),
\end{equation}
where significantly the response is given by the \emph{nonequilibrium} covariance between the observable and the conjugate coordinate, ${\rm Cov}(Q,V)=\langle QV\rangle-\langle Q\rangle\langle V\rangle$.
This result demonstrates that for a class of perturbations---where $U$ and $\mu$ are varied in unison---the FDT holds in its equilibrium form, arbitrarily far from equilibrium. 
That an equilibrium-like FDT held for certain time-dependent perturbations of diffusion processes was previously observed by Graham~\cite{Graham1977}.  Recently, we have extended this observation to arbitrary Markov processes~\cite{Chun2021}.
The value in rederiving this static response formula here is that it highlights its role as an important component of a more general framework for analyzing nonequilibrium response.

\emph{Energy perturbations.}\ Changes in the energy function $U$ represent a customary perturbation applied to probe a system's steady state.  While it can be challenging to interpret expressions for the response in this case, we can combine the predictions in \eqref{eq:Ebound} and \eqref{eq:Bbound1} to find simple thermodynamic constraints.

\begin{figure}[t]
\begin{center}
\includegraphics[width=.45\textwidth]{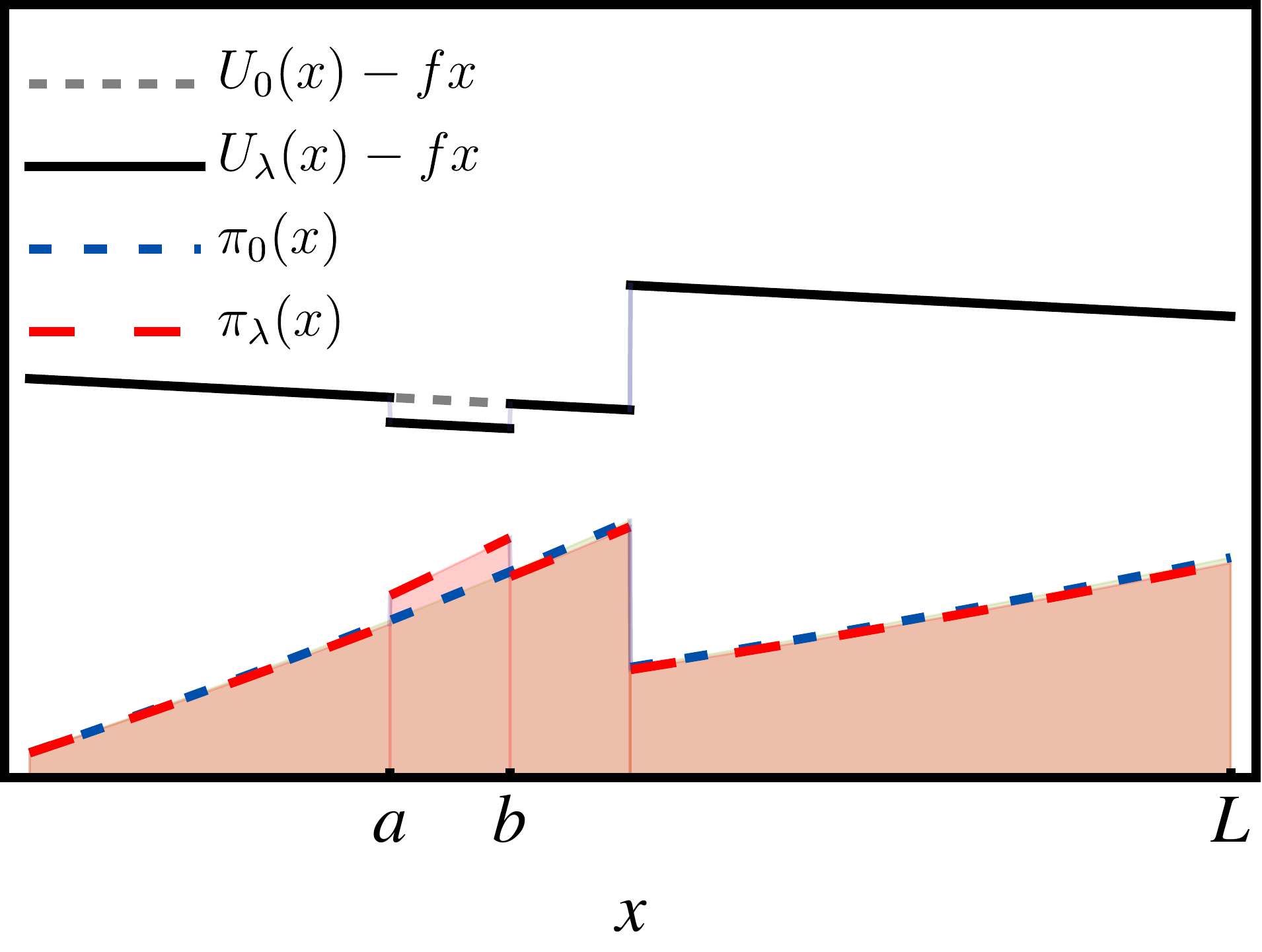}
\caption{Example of perturbing the energy landscape: Pictured is the ``effective potential''  as a function of position $x$ before the perturbation $U_0(x)-fx$ (gray dashed) and after lowering the energy in the region $x\in[a,b]$ by $\lambda I_{[a,b]}(x)$ (black).  This shifts the steady state distribution $\pi(x)$ from the orange dotted curve to the red long-dashed curve.}
\label{fig:energy_landscape_example}
\end{center}
\end{figure}

To apply our results, we have to focus on a perturbation where we shift the energy uniformly on a fixed interval $x\in[a,b]$ (Fig.~\ref{fig:energy_landscape_example}): specifically, $U_\lambda(x)= U(x)-\lambda I_{[a,b]}(x)$, where $I_{A}(z)$ is the indicator function taking the value $1$ when $z$ is in the set $A$ and $0$ otherwise.
 Our question is then how thermodynamics constrains the nonequilibrium response $R^{\rm neq}_{Q,U}=\partial_\lambda \langle Q\rangle=-\int_a^b\delta\langle Q\rangle/\delta U(z)dz$ of a (nonnegative) observable $Q$ to perturbations in $U$ with fixed thermodynamic driving ${\mathcal F}$.
Before addressing this question, however, let us first remind ourselves what a naive application of the FDT would have predicted, namely that the response would be given by the covariance between the observable $ Q(x)$ and the conjugate coordinate $I_{[a,b]}(x)$ as $R_{Q,U}^{\rm eq}={\rm Cov}\left(Q,I_{[a,b]}\right)$.

Now, let us proceed with perturbations of a nonequilibrium steady state (${\mathcal F}\neq 0$).
Observe that $U$ perturbations can be built from the sum
\begin{equation}
 \begin{split}
 R^{\rm neq}_{Q,U}&=-\int_a^b  \frac{\delta\langle Q\rangle}{\delta U(z)}dz\\
&=-\int_a^b \left[\frac{\delta \langle Q\rangle}{\delta U(z)}+\frac{\delta\langle Q\rangle}{\delta \ln \mu(z)}\right]-\frac{\delta\langle Q\rangle}{\delta \ln \mu(z)}dz.
\end{split}
\end{equation}
The first term is our coupled $\mu$-$U$ perturbation \eqref{eq:FDT1} that satisfies an equilibrium-like FDT \eqref{eq:FDT2} and is therefore equal to the covariance between the observable $Q$ and the conjugate coordinate $I_{[a,b]}$, ${\rm Cov}\left(Q,I_{[a,b]}\right)$, which is exactly the same as our naive prediction for the equilibrium response $R_{Q,U}^{\rm eq}$.
The remaining contribution can be constrained by the thermodynamic force using \eqref{eq:Bbound1} with the choices $Q_1(x)=Q(x)$ and $Q_2(x)=1$,
\begin{align}
\big|R^{\rm neq}_{Q,U}-R^{\rm eq}_{Q,U}\big|&=\left|\langle Q\rangle \int_a^b \frac{\delta\ln\langle Q\rangle}{\delta \ln \mu(z)}dz\right|\nonumber
\\
\label{eq:Ubound}
&\le \langle Q\rangle \tanh(|\mathcal{F}|/4).
\end{align}
The farther the system is from equilibrium, as measured by the force ${\mathcal F}$, the larger the possible nonequilibrium response. Alternatively, since $R^{\rm eq}_{Q,U}$ is the naive prediction from the FDT, we can interpret \eqref{eq:Ubound} as a quantitative bound on the violation of the FDT in terms of the nonequilibrium driving.

To illustrate this prediction, we analyzed the response of the steady-state density $\pi(x)$ itself, corresponding to the observable $Q(z;x)=\delta(z-x)$.
Denoting this response with a slight abuse of notation as $R^{\rm neq}_{x,U}$, the operative form of \eqref{eq:Ubound} is
\begin{equation}
\big|R^{\rm neq}_{x,U}-R^{\rm eq}_{x,U}\big|\le \pi(x)\tanh(|{\mathcal F}|/4)
\end{equation}
We choose perturbations of the energy landscape of the form $U(x)=U_0\Theta(x-L/2)$  where $\Theta(x-L/2)$ is the Heaviside step function and $U_0\in\{1,2,3\}$ is a constant (Fig.~\ref{fig:energy_landscape_example}). We further fix the mobility $\mu(x)=1$ and set the circumference of the circle to $L=1$. We numerically evaluated the response $R^{\rm neq}_{x,U}$ to energy perturbations on the interval $[x,b]$ as a function of ${\mathcal F}=f$  for 100 combinations of  $x$ and $b$ each sampled uniformly on the unit interval $[0,1]$. 
We have chosen the observation position $x$ to be on the edge of the perturbation region in order to enhance the sampling of highly responsive scenarios.
\begin{figure}[t]
\begin{center}
\includegraphics[width=.45\textwidth]{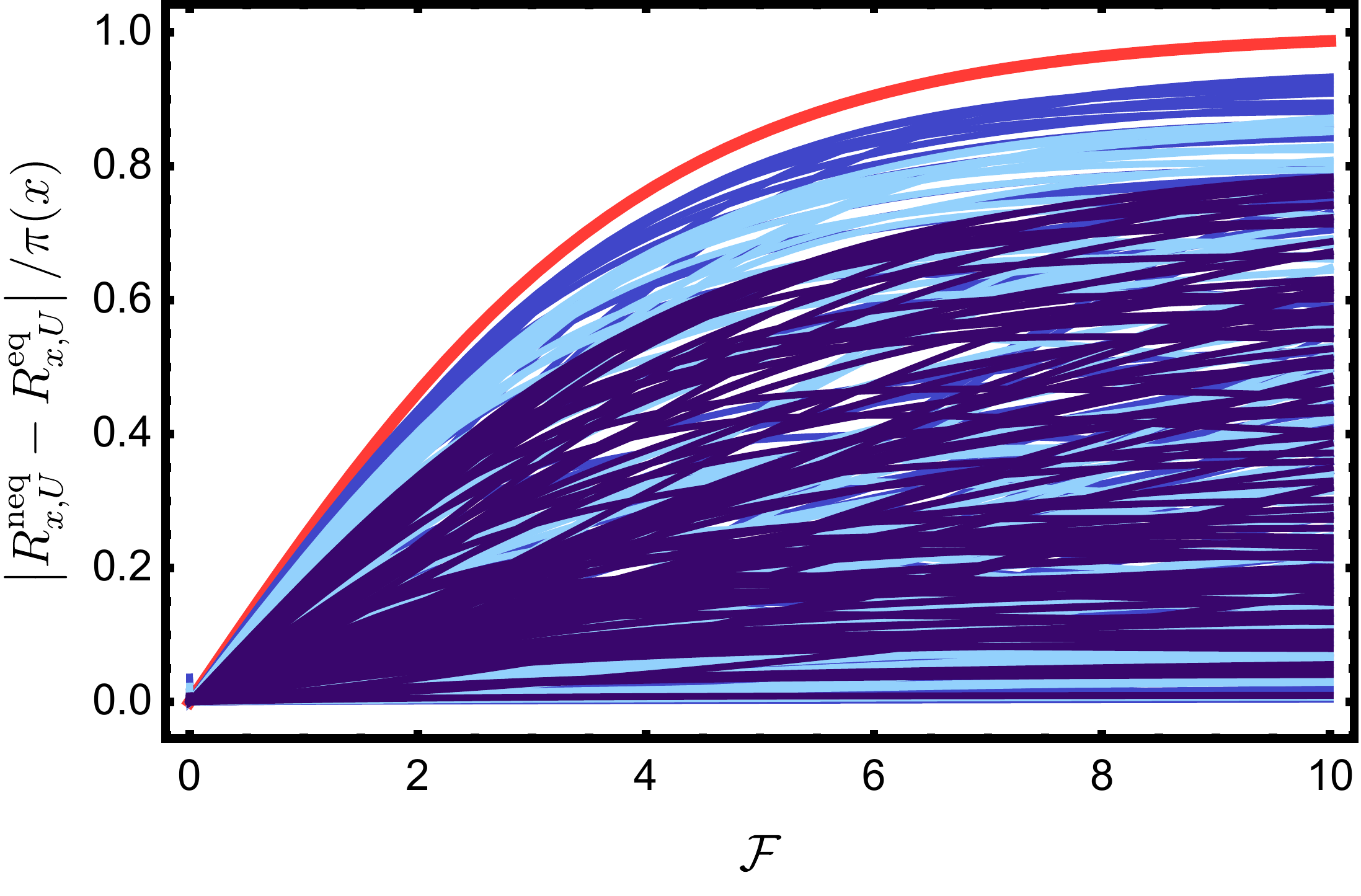}
\caption{Illustration of energy-perturbation thermodynamic bound: Normalized deviation of nonequilibrium response $\big|R^{\rm neq}_{x,U}-R^{\rm eq}_{x,U}\big|/\pi(x)$ at position $x$ for energy perturbations on the interval $[a,b]$ from an energy landscape $U(x)=U_0\Theta(x-L/2)$ given by the Heaviside step function multiplied by $U_0=1$ (dark blue), $2$ (light blue), $3$ (blue). Each color contains 100 randomly sampled pairs $(x=a)$ on the unit square. All curves fall below the predicted bound $\tanh(|{\mathcal F}|/4)$ (red line).  Other parameters: $L=1$ and $\mu(x)=1$.}
\label{fig:ln_mu_rho_response_example}
\end{center}
\end{figure}
The results presented in Fig.~\ref{fig:ln_mu_rho_response_example} verify that for all sampled parameter combinations the normalized deviation $\big|R^{\rm neq}_{x,U}-R^{\rm eq}_{x,U}\big|/\pi(x)$ remains below the predicted bound $\tanh(|{\mathcal F}|/4)$.

\emph{Discussion.}\ We have observed that any perturbation of a one-dimensional diffusion can be broken up into a linear combination of three types, which we term energy-mobility perturbations, mobility perturbations, and force perturbations.
For each class, we have derived either an equality or inequality characterizing the response in terms of the strength of the nonequilibrium driving.
One could have arrived at these predictions by discretizing the diffusion process and then using the bounds for discrete Markov dynamics reported previously in \cite{PhysRevX.10.011066}.
For completeness, we carry out this program explicitly in \cite{SM}, but note here that it requires a careful analysis of the limiting procedure.  
In light of this, our self-contained analysis based on the Fokker-Planck equation offers a more direct approach.

At the moment the analysis is limited in a handful of important ways.  
Our current methodology only works for one-dimensional systems, since  it is based on examining the analytic solution for the steady-state distribution, which is not known for higher-dimensional systems.
Moreover, discretizing higher-dimensional diffusions and then using the bounds reported in \cite{PhysRevX.10.011066} will not help either.  
We have checked that those inequalities are not sufficiently strong to provide useful limitations~\cite{SM}.
Even still, our results are suggestive that there is some thermodynamic structure in the nonequilibrium response of higher-dimensional diffusions, but it still remains to be investigated.

We have also limited our discussion to the response of state observables $Q(x)$ that are functions only of the system's position. The response of current observables, such as the velocity of the system, is an important extension of the current approach. Earlier studies on the Einstein relation connecting the velocity response (mobility) and diffusion coefficient for diffusive nonequilibrium steady states have also revealed FDT-like inequalities~\cite{doi:10.1098/rspa.2011.0046,Dechant_2018,Dechant6430}. Together these predictions suggest that there are also quantitative bounds on the response of generic current observables in terms of the thermodynamic force.

\bibliographystyle{apsrev}
\bibliography{Bib.bib}

\end{document}


\title{Supplemental Material for ``Thermodynamic constraints on the nonequilibrium response of one-dimensional diffusions"}
\maketitle 
\date{\today}
In this Supplemental Material, we first derive the transformation of perturbation formulas (3)-(5) [main text]. Then we derive our main predictions (6)-(8) [main text] quantifying the thermodynamic limits to steady-state response. Lastly, we demonstrate how our predictions could be obtained from the results in Ref.~\cite{PhysRevX.10.011066} as the limit of a discrete Markov jump process.

\section{Setup}
We have in mind a single degree of freedom with position $x$ diffusing on a circle of length $L$. Its dynamics as captured by the Fokker-Planck equation [(1), main text] are determined by two periodic functions $A(x)$ and $B(x)$. Our central observation is that we can reparameterize these dynamics as
\begin{align}
A(x)&=\mu(x)(-U'(x)+f),\label{eq:Transformation_A}
\\
B(x)&=\mu(x).\label{eq:Transformation_B}
\end{align} 
 in terms of three new periodic functions, which we call the ``energy'' $U$, ``mobility'' $\mu$, and ``force'' $f$ in analogy with the equation of motion for a mesoscopic Brownian particle in a viscous fluid. In terms of the new parameters, $\mu$, $U$, and $f$, the system's steady-state distribution $\pi(x)$ is given by the solution of the equation
\begin{equation}
-\partial_x [\mu(x)(-U'(x)+f)\pi(x)]+\partial_ x\left [\mu(x)\partial_x\pi(x)\right]=0.
\end{equation}
Even for general choices of the dynamics, this one-dimensional equation with periodic boundary conditions can always be determined  as \cite{Gardiner}
\begin{equation}\label{eq:ss}
\pi(x)=\frac{e^{-U(x)+fx}}{\mathcal N}\left(\int_0^x e^{U(x')-fx'-fL-\ln\mu(x')} dx'+\int_x^L e^{U(x')-fx'-\ln\mu(x')}  dx' \right),
\end{equation}
where ${\mathcal N}$ is a normalization constant determined by the requirement  $\int_0^L\pi(y)dy=1$.

In the following, we will be differentiating the steady-state distribution with respect to various system parameters.  We have found that for organizing these derivatives it is convenient to introduce an auxiliary function via 
\begin{align}\label{PiSExpression}
\pi(x)=\frac{1}{\mathcal N}\int_0^L dx' {\mathcal S}(x',x),
\end{align}
where
\begin{align}\label{eq:S}
 &{\mathcal S}(x',x)=e^{U(x')-U(x)-f(x'-x)-\ln\mu(x')}\left[e^{-fL}\Theta(x-x')+\Theta(x'-x)\right],\\
 \label{eq:N}
 &\mathcal N=\int_0^L\int_0^L  {\mathcal S}(x',x)\ dx' dx,
\end{align}
and $\Theta(z)$ is the Heaviside step function that is one for $z>0$ and zero otherwise. The quantity ${\mathcal S}$ acts like a continuous version of the spanning trees that appear in the Matrix Tree Theorem~\cite{Aleandri2020}, which were also key ingredients to the proofs of the thermodynamic bounds in Ref.~\cite{PhysRevX.10.011066}.

\section{Transformation of perturbations}
We are interested in the response of observables.
Our assertion is that any response can be reformulated in terms of derivatives with respect to our parameters $\mu$, $U$, and $f$. To verify this, we begin by first deriving the mapping from $A$ and $B$  to $\mu$, $U$, and $f$.
We then use this mapping to decompose any perturbation into perturbations of $\mu$, $U$, and $f$, thereby recovering (3)-(5) [main text].

Let us first invert the parameterization in \eqref{eq:Transformation_A} and \eqref{eq:Transformation_B} to find $\mu$, $U$ and $f$ in terms of $A$ and $B$.
First, by definition $\mu(x)=B(x)$.
Next, we eliminate $\mu$ from the transformation equations, by taking the ratio of $A$ and $B$
\begin{align}\label{eq:ABRatio}
\frac{A(x)}{B(x)}=-U'(x)+f.
\end{align} 
Exploiting the periodicity of $U$, we can integrate to single out 
\begin{align}\label{eq:Transformation_f}
f = \frac{1}{L}\int_0^L  \frac{A(z)}{B(z)}d{z}.
\end{align} 
Substituting \eqref{eq:Transformation_f} back into \eqref{eq:ABRatio}, leads to a closed equation for $U$ whose solution is
\begin{align}\label{eq:Transformation_U}
U(x) = -\int_0^x  \frac{A(z)}{B(z)}d{z} + \frac{x}{L}\int_0^L  \frac{A(z)}{B(z)}d{z} +U(0),
\end{align} 
with $U(0)$ an undetermined constant that has no affect on the dynamics.

Now, imagine that we apply a small perturbation by changing an external control parameter $\lambda$.  The Fokker-Planck equation is then modified by a small amount as $A_\lambda(x)=A(x)+\lambda \delta A(x)$ and $B_\lambda(x)=B(x)+\lambda \delta B(x)$, where $\delta A(x)$ and $\delta B(x)$ are periodic functions conjugate to the perturbation.
The response of the steady state to the perturbation in $\lambda$ can then be expressed via the chain rule as a combination of $A$ and $B$ functional derivatives as
\begin{align}
\frac{\partial}{\partial\lambda}&=\int_0^L dz\ \frac{\partial A_\lambda(z)}{\partial\lambda}\frac{\delta}{\delta A(z)}+\int_0^L dz\  \frac{\partial B_\lambda(z)}{\partial\lambda}\frac{\delta}{\delta B(z)}\\
& =\int_0^L dz\ \delta A (z)\frac{\delta}{\delta A(z)}+\int_0^L dz\  \delta B(z)\frac{\delta}{\delta B(z)}.\label{eq:lambda_der}
\end{align}
Using the mapping in \eqref{eq:Transformation_B}, \eqref{eq:Transformation_f}, and \eqref{eq:Transformation_U}, we can convert the functional derivatives with respect to $A$ and $B$ via the chain rule as
\begin{align}
\frac{\delta}{\delta A(z)}&=\int_0^Lds\ \frac{\delta U(s)}{\delta A(z)}\frac{\delta }{\delta U(s)} +\int_0^Lds\ \frac{\delta \mu(s)}{\delta A(z)}\frac{\delta }{\delta \mu(s)} +\frac{\delta f}{\delta A(z)}\frac{\partial }{\partial f}\\
&=\int_0^Lds\left(-\frac{\Theta(s-z)}{B(z)}+\frac{s}{L}\frac{1}{B(z)}\right)\frac{\delta}{\delta U(s)} + \frac{1}{L}\frac{1}{B(z)}\frac{\partial }{\partial f},
\end{align}
and
\begin{align}
\frac{\delta}{\delta B(z)}&=\int_0^Lds\ \frac{\delta U(s)}{\delta B(z)}\frac{\delta }{\delta U(s)} +\int_0^Lds\ \frac{\delta \mu(s)}{\delta B(z)}\frac{\delta }{\delta \mu(s)} +\frac{\delta f}{\delta B(z)}\frac{\partial }{\partial f}\\
&=\int_0^Lds\left(\Theta(s-z)\frac{A(z)}{B(z)^2}-\frac{s}{L}\frac{A(z)}{B(z)^2}\right)\frac{\delta}{\delta U(s)} +\frac{\delta}{\delta \mu(z)}- \frac{1}{L}\frac{A(z)}{B(z)^2}\frac{\partial }{\partial f}.
\end{align}
Substituting these expressions into the $\lambda$-derivative \eqref{eq:lambda_der}, we find after simplifying the integrals
\begin{equation}
\frac{\partial}{\partial\lambda} = \int_0^L ds\ \delta U(s) \frac{\delta}{\delta U(s)}+\int_0^L ds\ \delta\mu(s) \frac{\delta}{\delta \mu(s)}+\delta f\frac{\partial}{\partial f},
\end{equation}
where
\begin{align}
&\delta \mu(s) = \delta B(s)\\
&\delta U(s)=-\int_0^s dz\ \left(\frac{\delta A(z) B(z)-A(z)\delta B(z)}{B(z)^2}\right)+\frac{s}{L}\int_0^Ldz\ \left(\frac{\delta A(z) B(z)-A(z)\delta B(z)}{B(z)^2}\right)\\
&\delta f = \frac{1}{L} \int_0^L dz\  \left(\frac{\delta A(z) B(z)-A(z)\delta B(z)}{B(z)^2}\right),
\end{align}
recapitulating the functional differentials found  in (3)-(5) [main text]. Furthermore, we do not study $\mu$, $U$, and $f$ perturbations explicitly, but instead constrain linear combinations of these perturbations. Nevertheless, we can express the response in terms of these linear combinations as 
\begin{equation}
\frac{\partial}{\partial\lambda} = \int_0^L ds\ \delta U(s) \left(\frac{\delta}{\delta U(s)} +\frac{\delta}{\delta \ln \mu(s)}\right) +\int_0^L ds\ \big(\mu(s) \delta\mu(s) -\delta U(s)\big)\frac{\delta}{\delta \ln \mu(s)}+\delta f\frac{\partial}{\partial f}
\end{equation}

\section{Energy-mobility perturbations}
Our first result in (6) [main text] is an equality for the response of the steady-state average of an observable $Q(x)$ to a coordinated perturbation in the energy and mobility: $(\delta/\delta U(z)+\delta/\delta\ln\mu(z))\langle Q\rangle$. 

Let us proceed by first analyzing the derivatives on $\mathcal S(x',x)$ and $\mathcal N$: 
\begin{align}
&\frac{\delta{\mathcal S}(x',x)}{\delta U(z)}+\frac{\delta {\mathcal S}(x',x)}{\delta\ln\mu(z)}= -\delta(x-z) {\mathcal S}(x',x),\qquad \frac{\delta{\mathcal N}}{\delta U(z)}+\frac{\delta{\mathcal N}}{\delta\ln\mu(z)}=-{\mathcal N}\pi(z).
\end{align} 
Using these expressions, we then find for the derivative of $\pi(x)$,

\begin{align}
\frac{\delta \pi(x)}{\delta U(z)}+\frac{\delta\pi(x)}{\delta \ln \mu(z)}&=-\frac{1}{\mathcal N}\int_0^L dx' \delta(x-z){\mathcal S}(x',x)-\frac{1}{{\mathcal N}^2}\left[-{\mathcal N}\pi(z)\right]\int_0^L dx' {\mathcal S}(x',x)\\
&=-\pi(x)(\delta(x-z)-\pi(z)).
\end{align} 
From this expression, we readily obtain equation (6) [main text] for the response of an observable $\langle Q\rangle = \int_0^L Q(x)\pi(x)dx$ as
\begin{equation}
\frac{\delta \langle Q \rangle}{\delta U(z)}+\frac{\delta\langle Q \rangle}{\delta \ln \mu(z)}=\int_0^L Q(x)\left[\frac{\delta \pi(x)}{\delta U(z)}+\frac{\delta\pi(x)}{\delta \ln \mu(z)}\right]dx=-\pi(z)(Q(z)-\langle Q \rangle).
\end{equation} 

\section{Mobility perturbations}
Our next prediction, (7) [main text], is a thermodynamic bound on perturbations with respect to the mobility $\mu(x)$:  
\begin{equation}\label{eq:mobility1}
\left|\int_a^b\frac{\delta\ln(\langle Q_1\rangle/\langle Q_2\rangle)}{\delta\ln\mu(z)}dz\right|\le\tanh(\left|f\right| L/4),
\end{equation}
where $Q_1(x)$, $Q_2(x)$ are bounded observables, and their values depend only on position $x$. Without loss of generality, we can assume that they are non-negative ($Q(x)\ge 0$), because we can always redefine them by subtracting off the minimum ($Q'(x)=Q(x)-Q_{\rm min}$). 

To evaluate the derivative in \eqref{eq:mobility1}, we proceed by first differentiating ${\mathcal S}$ and ${\mathcal N}$:
\begin{equation}
\frac{\delta {\mathcal S}(x',x)}{\delta\ln\mu(z)}=-\delta(x'-z){\mathcal S}(x',x),\qquad \frac{\delta {\mathcal N}}{\delta\ln\mu(z)}=-\int_0^Ldy\ {\mathcal S}(z,y).
\end{equation}
Consequently,
\begin{equation}
\frac{\delta \pi(x)}{\delta\ln\mu(z)} = -\frac{1}{{\mathcal N}}{\mathcal S}(z,x)+\frac{\pi(x)}{\mathcal N}\int_0^Ldy\ {\mathcal S}(z,y).
\end{equation}
With this expression, we can readily obtain an expression for the response of an observable as
\begin{align}
\frac{\delta\langle Q\rangle}{\delta\ln\mu(z)} &= \int_0^LQ(x)\left[-\frac{1}{{\mathcal N}}{\mathcal S}(z,x)+\frac{\pi(x)}{\mathcal N}\int_0^Ldy\ {\mathcal S}(z,y)\right]dx\\
\label{eq:Qresponse}
&=-\frac{1}{{\mathcal N}}\int_0^LQ(x) {\mathcal S}(z,x)dx+\frac{\langle Q\rangle}{\mathcal N}\int_0^L\ {\mathcal S}(z,y) dy.
\end{align}
We are now in a position to evaluate the derivative in \eqref{eq:mobility1}:
\begin{align}
\int_a^b\frac{\delta\ln(\langle Q_1\rangle/\langle Q_2\rangle)}{\delta\ln\mu(z)} dz&=\int_a^b \frac{\frac{\delta\langle Q_1\rangle}{\delta\ln\mu(z)} \langle Q_2\rangle-\frac{\delta\langle Q_2\rangle}{\delta\ln\mu(z)}\langle Q_1\rangle }{\langle Q_1\rangle\langle Q_2\rangle}dz.
\end{align}
Upon substitution of \eqref{eq:Qresponse}, we see that the terms linear in average of the observable in \eqref{eq:Qresponse} cancel, leaving
\begin{align}\label{eq:Qresponseratio}
\int_a^b\frac{\delta\ln(\langle Q_1\rangle/\langle Q_2\rangle)}{\delta\ln\mu(z)} dz&= -\frac{1}{\mathcal N}\frac{\left(\int_a^b\int_0^L Q_1(x)S(z,x)dzdx\right) \langle Q_2\rangle-\left(\int_a^b\int_0^L Q_2(x)S(z,x)dzdx\right)\langle Q_1\rangle }{\langle Q_1\rangle\langle Q_2\rangle}
\end{align}
To simplify this expression, we note that average of any observable can also be expressed in terms of ${\mathcal S}$ as
\begin{equation}
\langle Q\rangle = \frac{1}{\mathcal N}\int_0^L\int_0^LQ(x) S(x',x)dx' dx.
\end{equation}
Upon substitution of this formula into \eqref{eq:Qresponseratio}, we find that the result can be conveniently expressed in terms of the integrals
\begin{align}
q_1&=\int_{z\in[a,b]}\int_0^L Q_1(x){\mathcal S}(z,x)\ dxdz
\\
{\bar q}_1&=\int_{z\notin[a,b]}\int_0^L Q_1(x){\mathcal S}(z,x)\ dxdz
\\
q_2&=\int_{z\in[a,b]}\int_0^L Q_2(x){\mathcal S}(z,x)\ dxdz
\\
{\bar q}_2&=\int_{z\notin[a,b]}\int_0^L Q_2(x){\mathcal S}(z,x)\ dxdz
\end{align}
as
\begin{equation}\label{eq:Qq}
\int_a^b\frac{\delta\ln(\langle Q_1\rangle/\langle Q_2\rangle)}{\delta\ln\mu(z)}dz=\frac{q_1 {\bar q}_2-{\bar q}_2 q_1}{({\bar q}_1+q_1)({\bar q}_2+q_2)}.
\end{equation}
The notation here is reminiscent of the derivation Ref.~\cite{PhysRevX.10.011066}, which will allow us to import those methods directly.

Noting that ${\bar q}_1$, $q_1$, ${\bar q}_2$, and $q_2$ are all non-negative, the denominator of \eqref{eq:Qq} is bounded by the inequality of arithmetic and geometric means:
\begin{equation}
({\bar q}_1+q_1)({\bar q}_2+q_2)={\bar q}_1 {\bar q}_2+{\bar q}_1 q_2+q_1 {\bar q}_2+q_1 q_2\ge q_1 {\bar q}_2+{\bar q}_1 q_2+2\sqrt{{\bar q}_1 {\bar q}_2 q_1 q_2}=(\sqrt{{\bar q}_1 q_2} +\sqrt{q_1 {\bar q}_2} )^2,
\end{equation}
where the equality is saturated when ${\bar q}_1 {\bar q}_2=q_1 q_2$. The numerator can also be factored
\begin{equation}
q_2 {\bar q}_1-q_1 {\bar q}_2= (\sqrt{{\bar q}_1 q_2}-\sqrt{q_1 {\bar q}_2})(\sqrt{{\bar q}_1 q_2} +\sqrt{q_1 {\bar q}_2} ).
\end{equation}
The result is
\begin{equation}
\left|\int_a^b\frac{\delta\ln(\langle Q_1\rangle/\langle Q_2\rangle)}{\delta\ln\mu(z)}dz\right|\le
\left|\frac{\sqrt{{\bar q}_1 q_2}-\sqrt{q_1 {\bar q}_2}}{\sqrt{{\bar q}_1 q_2} +\sqrt{q_1 {\bar q}_2}}\right|=
\tanh\left(\frac{1}{4}\left|\ln\frac{q_1 {\bar q}_2}{{\bar q}_1 q_2}\right|\right).
\end{equation}

Our last step is to bound the ratio $q_1 {\bar q}_2/{\bar q}_1 q_2$:
\begin{align}
\frac{q_1 {\bar q}_2}{{\bar q}_1 q_2}&=\frac{\int_0^L\int_0^L dx_1 dx_0\ \int_{z_1\in[a,b]}dz_1\int_{z_0\notin[a,b]}dz_0 \ Q_1(x_0) Q_2(x_1){\mathcal S}(z_1,x_0)\mathcal S(z_0,x_1)}{\int_0^L\int_0^L dx_1 dx_0\  \int_{z_1\in[a,b]}dz_1\int_{z_0\notin[a,b]}dz_0 \ Q_1(x_0) Q_2(x_1)\mathcal S(z_0,x_0)\mathcal S(z_1,x_1)}
\\
&=\frac{\int_0^L\int_0^L dx_1 dx_0 \int_{z_1\in[a,b]}dz_1\int_{z_0\notin[a,b]}dz_0 \ W(x_0,x_1,z_0,z_1)\frac{\mathcal S(z_1,x_0)\mathcal S(z_0,x_1)}{\mathcal S(z_0,x_0)\mathcal S(z_1,x_1)}}{\int_0^L\int_0^L dx_1 dx_0 \int_{z_1\in[a,b]}dz_1\int_{z_0\notin[a,b]}dz_0 \ W(x_0,x_1,z_0,z_1)},
\end{align}
where we introduced the four-dimensional non-negative weight function
\begin{equation}
W(x_0,x_1,z_0,z_1)=Q_1(x_0)Q_2(x_1)\mathcal S(z_0,x_0)\mathcal S(z_1,x_1)\ge 0.
\end{equation}
Therefore the ratio $q_1 {\bar q}_2/{\bar q}_1 q_2$ can be viewed as the weighted average of an observable ($\langle\cdot\rangle_W$), which we can bound by its maximum as
\begin{align}
\frac{q_1 {\bar q}_2}{{\bar q}_1 q_2}&=\left\langle\frac{\mathcal S(z_1,x_0)\mathcal S(z_0,x_1)}{\mathcal S(z_0,x_0)\mathcal S(z_1,x_1)}\right\rangle_W\\
&\le \max_{\{z_0,z_1,x_0,x_1\}}\frac{\mathcal S(z_1,x_0)\mathcal S(z_0,x_1)}{\mathcal S(z_0,x_0)\mathcal S(z_1,x_1)}
\\
&=\max_{\{z_0,z_1,x_0,x_1\}} \frac{\left[e^{-fL}\Theta(x_0-z_1)+\Theta(z_1-x_0)\right]\left[e^{-fL}\Theta(x_1-z_0)+\Theta(z_0-x_1)\right]}{\left[e^{-fL}\Theta(x_0-z_0)+\Theta(z_0-x_0)\right]\left[e^{-fL}\Theta(x_1-z_1)+\Theta(z_1-x_1))\right]}
\\
&=e^{\left|f\right| L},
\end{align} 
where the last equality holds when, for example, $f>0$ and $z_0>x_1>z_1>x_0$. Equation \eqref{eq:mobility1} follows immediately.
\\
\\
\section{Force perturbation}
The final prediction is a bound on the force response, (8) [main text], reproduced here in a slightly modified form 
\label{eq:bounds}
\begin{align}\label{eq:fbound1}
&\left|\frac{\partial \ln(\langle Q_1\rangle/\langle Q_2\rangle)}{\partial f}\right|\le L.
\end{align}

To organize the derivatives with respect to the force $f$, we will find it convenient to use the function
\begin{equation}
{\mathcal O}(x',x)=(x'-x+L)\Theta(x-x')+(x'-x)\Theta(x'-x),
\end{equation}
which we note for later use is bounded $0\le {\mathcal O}\le L$.
Then, we have
\begin{align}
\frac{{\partial\mathcal S}(x',x)}{\partial f}=-{\mathcal O}(x',x){\mathcal S}(x',x),\qquad\frac{{\partial\mathcal N}}{\partial f}=-\int_0^L\int_0^L dx'dx\ {\mathcal O}(x',x){\mathcal S}(x',x),
\end{align}
so that
\begin{equation}
\frac{\partial \pi(x)}{\partial f}=-\frac{1}{\mathcal N}\int_0^L dx' {\mathcal O}(x',x){\mathcal S}(x',x)+\frac{\pi(x)}{\mathcal N}\int_0^L\int_0^L dx'dx''\ {\mathcal O}(x',x''){\mathcal S}(x',x'').
\end{equation}
As a result the response of an observable can be expressed as
\begin{equation}\label{eq:Qf}
\frac{\partial\langle Q\rangle}{\partial f}=-\frac{1}{\mathcal N}\int_0^L \int_0^Ldx'dx\ Q(x){\mathcal O}(x',x){\mathcal S}(x',x)+\frac{\langle Q\rangle}{\mathcal N}\int_0^L\int_0^L dx'dx\ {\mathcal O}(x',x){\mathcal S}(x',x).
\end{equation}

With these formulas in hand, we can now address the derivative in \eqref{eq:fbound1}.
Upon substitution of \eqref{eq:Qf} into \eqref{eq:fbound1}, we find that the second terms in \eqref{eq:Qf} linear in the average of the observables cancel, resulting in the expression
\begin{align}\label{eq:Qfweight}
\left|\frac{\partial \ln(\langle Q_1\rangle/\langle Q_2\rangle)}{\partial f}\right|=\left|\frac{\int_0^L \int_0^Ldx'dx\ {\mathcal O}(x',x)Q_2(x){\mathcal S}(x',x)}{\int_0^L \int_0^Ldx'dx\ Q_2(x){\mathcal S}(x',x)}-\frac{\int_0^L \int_0^Ldx'dx\ {\mathcal O}(x',x)Q_1(x){\mathcal S}(x',x)}{\int_0^L \int_0^Ldx'dx\ Q_1(x){\mathcal S}(x',x)}\right|,
\end{align}
after simplification using the definition of ${\mathcal N}$.
A particularly useful interpretation presents itself after we note that $Q(x){\mathcal S}(x',x)\ge 0$.
Therefore each ratio can be interpreted as a normalized average of ${\mathcal O}$ with observable-dependent weight $Q(x){\mathcal S}(x',x)\ge 0$, which we denote as $\langle \cdot \rangle_{Q}$.
The result is that we can  express \eqref{eq:Qfweight} as
\begin{align}
\left|\frac{\partial \ln(\langle Q_1\rangle/\langle Q_2\rangle)}{\partial f}\right|=\big|\langle {\mathcal O}\rangle_{Q_2}-\langle {\mathcal O}\rangle_{Q_1}\big|\le L,
\end{align}
where the bound follows from $0\le {\mathcal O}\le L$, completing the derivation.

\section{Connection to Markov Jump Processes}

The aim of this section is to discuss the relationship between the present study and previous work on thermodynamic limitations to steady-state response in discrete Markov jump processes~\cite{PhysRevX.10.011066}.

\subsection{Review of thermodynamic limits to response for discrete Markov jump processes}
As we are only interested in stochastic processes on a ring, we will introduce the ideas and results from Ref.~\cite{PhysRevX.10.011066} specialized to this context.

We have in mind a system of $N$ discrete states at positions $x_i=i\Delta x$ around a ring of length $L=N\Delta x$.
We label these states as of $i=0,\dots, N$, where we identify the redundant state $i=N$ with $i=0$ to enforce the periodic boundary conditions.
The probability to find the system in state $i$ at time $t$ is then governed by the Master equation~\cite{Gardiner}
\begin{equation}\label{eq:Master}
\dot p_i(t)=\sum_{j=0}^{N-1}W_{ij}p_j(t),
\end{equation}
where the off-diagonal entries of the transition rate matrix $W_{ij}$ specify the probability per unit time to jump from state $j$ to state $i$, and $W_{ii}=-\sum_{j\neq i} W_{ji}$. 
As only nearest-neighbor hops are allowed, the only nonzero transition rates are those for which $i$ and $j$ differ by one; thus, all rates are of the form $W_{i+1,i}$ or $W_{i-1,i}$, corresponding to `right' and `left' hops.
As the state space is irreducible, a unique stationary distribution $\pi_i$ exists and can be obtained as the solution of
\begin{equation}
\sum_{j=0}^{N-1}W_{ij}\pi_j=0.
\end{equation} 
Thermodynamics is included in the model by identifying the log-ratio of rates around cycles as the thermodynamic force driving the system out of equilibrium.
As a ring only has a single cycle, the sole thermodynamic force is
\begin{equation}\label{eq:ThermoForce}
F_{C}=\ln\frac{W_{0,N-1}\cdots W_{2,1}W_{1,0}}{W_{N-1,0}\cdots W_{1,2}W_{0,1}}.
\end{equation}

Reference~\cite{PhysRevX.10.011066} introduced a parameterization of the transition rate matrix in terms of vertex parameters $E_i$, symmetric edge parameters $B_{i+1,i}=B_{i,i+1}$, and asymmetric edge parameters $F_{i+1,i}=-F_{i,i+1}$:
\begin{align}\label{eq:rates}
W_{i+1,i} =e^{-(B_{i+1,i}-E_i-F_{i+1,i}/2)}, \qquad W_{i-1,i} =e^{-(B_{i,i-1}-E_i+F_{i,i-1}/2)}.
\end{align}
Nonequilibrium effects are included in this parameterization solely through the asymmetric edge parameters, which can be seen by substituting this decomposition into the definition of the thermodynamic force~\eqref{eq:ThermoForce} to conclude
\begin{equation}\label{eq:ThermoForce2}
F_{C}=\sum_{i=0}^{N-1} F_{i+1,i}.
\end{equation}

The main predictions of Ref.~\cite{PhysRevX.10.011066} are then a series of equalities and inequalities for the derivative of the steady state distribution with respect to these three parameter families.
Here, we present forms most relevant for our present discussion.

\emph{Vertex parameters}: An equality for vertex parameter perturbations can be obtained from Eq.~(13) of Ref.~\cite{PhysRevX.10.011066},
\begin{align}
\label{PRXEBound}
\sum_{j=0}^{N-1} V_j\frac{\partial\langle Q\rangle}{\partial E_j}=-\sum_{j=0}^{N-1} V_j\pi_j(Q_j-\langle Q\rangle)=-{\rm Cov}(Q,V),
\end{align}
where the state space observable $V_i$ is the conjugate coordinate to the perturbation.

\emph{Symmetric edge parameters}: 
If we perturb the $B_{i+1,i}$ of all the edges between a pair of nodes at positions $x_a=a\Delta x$ and $x_b=b\Delta x$ then Eq.~(20) of Ref.~\cite{PhysRevX.10.011066} predicts
\begin{align}
\label{PRXBBound}
\left|\sum_{i=a}^{b-1}\frac{\partial\ln(\langle Q_1 \rangle/\langle Q_2 \rangle)}{\partial B_{i+1,i}}\right|&\le\tanh(\left| F_C\right|/4).
\end{align}

\emph{Asymmetric edge parameters}: 
For perturbations of all the $F_{i+1,i}$ all the way around the ring, one can deduce from Eq.~(21) of Ref.~\cite{PhysRevX.10.011066} using techniques in that paper an equality of the form,
\begin{align}
\label{PRXFBound}
\left|\sum_{i=0}^{N-1}\frac{\partial \ln(\langle Q_1\rangle/\langle Q_2\rangle)}{\partial F_{i+1,i}}\right|&\le N,
\end{align}
although this expression does not explicitly appear.

It is the continuous limits of these formulas that are operative for diffusion processes.
To make this connection, we first have to develop the mapping between this discrete Markov jump process and its limit as a continuous diffusion process, obtained as the spacing between lattice points tends to zero, $\Delta x\to 0$.

\subsection{Discrete approximation of a continuous diffusion process} 

Motivated by the structure of the decomposition of the transition rate matrix in \eqref{eq:rates}, we now look to construct a Markov jump process that has a well defined continuous limit as a diffusion process, and that maintains that structure.

To begin, we first introduce a smooth probability density $\rho(x,t)$ such that the probability the system is between $x_i-\Delta x/2$ and $x_{i}+\Delta x/2$ at time $t$ is given by $\rho(x_i,t)=p_i(t)/\Delta x$.
We also introduce three more smooth functions of space $E(x)$, $B(x)$ and $f$ such that
\begin{align}
&E(x_i) = E_i\\
&B(x_i)=B_{i+1,i}=B_{i,i+1}\\
&f \Delta x= F_{i+1,i}=-F_{i,i+1}.
\end{align}
Notice that we have assigned the ``location'' of $B_{i+1,i}$ to the position with the smaller index. 
In addition, to have a well-defined limit the asymmetric edge parameters need to be linear in $\Delta x$, and a constant value is sufficient to include all possible nonconservative effects.
In terms of these functions, the transition rates \eqref{eq:rates} become
\begin{align}
&W_{+}(x_i)\equiv W_{i+1,i} =e^{-(B(x_i)-E(x_i)-f\Delta x/2)}\\
&W_{-}(x_i)\equiv W_{i-1,i} =e^{-(B(x_{i-1})-E(x_i)+f\Delta x/2)}=e^{-(B(x_{i}-\Delta x)-E(x_i)+f\Delta x/2)},
\end{align}
and thermodynamic force \eqref{eq:ThermoForce} simplifies to $F_C = \sum_{i=0}^{N-1} f\Delta x= fL$.

With this setup the procedure to carry out the limit $\Delta x\to 0$ is as follows: We substitute these definitions into the Master equation \eqref{eq:Master}, expand for small $\Delta x$, and then diffusively rescale time $t\to t/\Delta x^2$. The result is the Fokker-Planck equation
\begin{align}
\partial_{t}\rho(x,t)&=-\partial_x\left[e^{-{\bm (}B(x)-E(x){\bm )}}(E'(x)-f) \rho(x,t)\right]+\partial_x\left[e^{-{\bm (}B(x)-E(x){\bm )}}\partial_x \rho(x,t)\right].
\end{align}
This is of the form in (1) [main text]. 
Codifying the observation that interesting results in the discrete case correspond to separate perturbations in the $E$, $B$, and $f$ functions, then suggests the decomposition introduced in (2) [main text] via the identification
\begin{equation}\label{eq:map}
\mu(x)=\exp{\bm (}E(x)-B(x){\bm)},\quad U(x)=E(x).
\end{equation}

\subsection{Diffusion limits of thermodynamic bounds}
Having established a consistent discretization of our diffusion process, we turn to utilizing the thermodynamic bounds for discrete Markov processes \eqref{PRXEBound} - \eqref{PRXFBound} to prove the analogous thermodynamic bounds for the continuous limit.

In make this connection, we will repeatedly face the situation where we have to convert a derivative with respect to a finite collection of variables, like the $\{E_i\}$ or $\{B_{i+1,i}\}$, into a functional derivative as the spacing tends to zero ($\Delta x\to 0$).
In preparation for these calculations, we first present this relationship in general and then exploit it in the following.
To this end, let us consider two smooth functions $f(x)$ and $g(x)$, and the functional ${\mathcal I}[f]$.
In the discrete picture, we only evaluate these functions at the positions $x_i$, with values $f(x_i)$ and $g(x_i)$.
The functional is then a function ${\mathcal I}{\bm (}\{f(x_i)\}{\bm )}$ of the finite set of values $\{f(x_i)\}$, but is assumed to tend smoothly to ${\mathcal I}{\bm (}\{f(x_i)\}{\bm )}\to {\mathcal I}[f]$ as $\Delta x\to 0$.
With this setup, as $\Delta x\to 0$  the definitions of the derivative and functional derivative are connected by 

\begin{align}\label{eq:funcDer}
\lim_{\Delta x\to0}\sum_{i=0}^{N-1} g(x_i)\frac{\partial{\mathcal I}{\bm (}\{f(x_i)\}{\bm )}}{\partial f(x_i)}=\int_0^Lg(x)\frac{\delta{\mathcal I}[f]}{\delta f(x)}dx.
\end{align}
Let us now address each type of perturbation in turn.

\emph{Vertex parameters}:  For the vertex derivatives, we first replace $E_i=E(x_i)$ and $V_i=V(x_i)$, and then take the continuous limit
\begin{equation}
\lim_{\Delta x\to 0}\sum_{j=0}^{N-1}  V_j\frac{\partial\langle Q\rangle}{\partial E_j}=\lim_{\Delta x\to 0}\sum_{j=0}^{N-1} V(x_j)\frac{\partial\langle Q\rangle}{\partial E(x_j)}=\int_0^L V(x)\frac{\delta \langle Q\rangle}{\delta E(x)}dx,
\end{equation}
where we used \eqref{eq:funcDer} with ${\mathcal I}=\langle Q\rangle$, $f(x)=E(x)$, and $g(x)=V(x)$. 
Inserting this expression into \eqref{PRXEBound} and applying the identification $U(x)=E(x)$~\eqref{eq:map}, we arrive at expression equivalent to Eq.~(14) [main text].

\emph{Symmetric edge parameters}: 
When we perturb all the symmetric edge parameters between positions $x_a=a \Delta x$ and $x_b=b \Delta x$, we obtain the response in the continuous limit by first replacing $B_{i+1,i}=B(x_i)$, and then 
\begin{align}
\lim_{\Delta x\to 0}\sum_{i=a}^{b-1}\frac{\partial\ln(\langle Q_1 \rangle/\langle Q_2 \rangle)}{\partial B_{i+1,i}}=
\lim_{\Delta x\to 0}\sum_{i=a}^{b-1}\frac{\partial\ln(\langle Q_1 \rangle/\langle Q_2 \rangle)}{\partial B(x_i)}=
\int_{x_a}^{x_b} \frac{\delta\ln(\langle Q_1 \rangle/\langle Q_2 \rangle)}{\delta B(x)}dx,
\end{align}
where we have utilized \eqref{eq:funcDer}, with ${\mathcal I}=\ln(\langle Q_1 \rangle/\langle Q_2 \rangle)$, $f(x)=B(x)$, and $g(x)=I_{[a,b]}(x)$ is the indicator function on the set $x\in [a,b]$. 
Substituting into \eqref{PRXBBound}, noting the change of variables $\delta\ln\mu(x)=-\delta B(x)$ (with $E(x)$ fixed) from \eqref{eq:map}, and that the sole thermodynamic force is $F_C=fL$ we arrive at Eq.~(7) [main text].

\emph{Aymmetric edge parameters}: Lastly, for asymmetric edge perturbations, we link the $f$-perturbations via
\begin{equation}
\left|\frac{\partial \ln(\langle Q_1\rangle/\langle Q_2\rangle)}{\partial f}\right|=\left|\lim_{\Delta x\rightarrow 0}\sum_{i=1}^N\frac{\partial \ln(\langle Q_1\rangle/\langle Q_2\rangle)}{\partial (F_{i+1,i}/\Delta x)}\right|\le \lim_{\Delta x \to 0} N\Delta x=L,
\end{equation}
where the inequality is due to \eqref{PRXFBound}, and the desired result Eq.~(8) [main text] follows. 

\subsection{Failure of bounds in the continuous limit for higher dimensions}

It turns out that the results known for discrete Markov process~\cite{PhysRevX.10.011066} are not sufficient to constrain the steady-state response of diffusion processes in higher dimensions.

To demonstrate this possibility, we focus here on a two-dimensional diffusion process with positions $(x,y)$ on a torus whose circumferences in both directions are $L$.
As before, we discretize the dynamics by placing the evolution on a square lattice with lattice spacing $l$, and discretized positions $(x_i,y_j)=(il,jl)$.
The transition rates are only nonzero for nearest neighbor hops in the positive and negative $x$ and $y$ directions.
Motivated by our previous discussion we introduce the smooth functions defined on the torus, $B_x(x,y)$,  $B_y(x,y)$, $E(x,y)$, $f_x$ and $f_y$, allowing us to specify the transition rates
\begin{align}
W_{i+1,i}^{j} &= e^{-{\bm (}B_x(x_i,y_j)-E(x_i,y_j)-f_x l/ 2{\bm )})}\\
W_{i-1,i}^{j} &= e^{-{\bm (}B_x(x_{i-1},y_j)-E(x_i,y_j)+f_x l/ 2{\bm )}}\\
W_i^{j+1,j} &= e^{-{\bm (}B_y(x_i,y_j)-E(x_i,y_j)-f_y l/ 2{\bm )}}\\
W_i^{j-1,j} &= e^{-{\bm (}B_y(x_i,y_{j-1})-E(x_i,y_j)+f_y l/ 2{\bm )}}.
\end{align}
For similar reasons as above, these rates limit to a diffusion process as $l\to 0$

Now imagine we perturb all the symmetric edge parameters in a square region from $x_a=a l$ to $x_b=b l$ and from $y_{a'}=a' l$ to $y_{b'}=b' l$, totaling $N_{\rm e} = (b-a)(b'-a'-1)+(b-a-1)(b'-a')$ edges.
Now Eq.~(20) of Ref.~\cite{PhysRevX.10.011066} predicts that the response is no worse than the number of {\it vertices on the perimeter} of this region $N_{\rm p}=2(b-a+b'-a')$ as
\begin{equation}
\left|\sum_{i=a}^{b-1}\sum_{j=a'}^{b'-1}\left(\frac{\partial\ln(\langle Q_1\rangle/\langle Q_2\rangle)}{\partial B_x(x_i,y_j)}+\frac{\partial\ln(\langle Q_1\rangle/\langle Q_2\rangle)}{\partial B_y(x_i,y_j)}\right)\right|\le N_{\rm p}-1.
\end{equation}
In the continuous limit $l\to0$, the left hand side tends to a finite value given by the functional derivative
\begin{equation}
\lim_{l\to 0}\sum_{i=a}^{b-1}\sum_{j=a'}^{b'-1}\left(\frac{\partial\ln(\langle Q_1\rangle/\langle Q_2\rangle)}{\partial B_x(x_i,y_j)}+\frac{\partial\ln(\langle Q_1\rangle/\langle Q_2\rangle)}{\partial B_y(x_i,y_j)}\right)=
\int_{x_a}^{x_b}\int_{y_{a'}}^{y_{b'}} \frac{\delta\ln(\langle Q_1\rangle/\langle Q_2\rangle)}{\delta B_x(x,y)}+\frac{\delta\ln(\langle Q_1\rangle/\langle Q_2\rangle)}{\delta B_y(x,y)}dy dx.
\end{equation}
However, the right hand side tends to infinity, since the number of vertices on the perimeter grows without bound as the spacing tends to zero.
Thus, the inequalities derived in Ref.~\cite{PhysRevX.10.011066}  for discrete Markov processes are uninformative in the continuous limit in dimensions above one.

\bibliographystyle{apsrev}
\bibliography{Bib.bib}